\documentclass{sig-alternate}
\usepackage[english]{babel}
\usepackage[utf8x]{inputenc}

\usepackage[]{subfigure}

\usepackage[final]{listings}
\usepackage{float}

\usepackage{tikz}
\usepackage{pgfplots}
\usetikzlibrary{calc}
\usetikzlibrary{backgrounds}
\usepgflibrary{shapes}
\usetikzlibrary{through}
\usetikzlibrary{intersections}

\usepackage{xifthen}

\usepackage{hyperref}
\usepackage{esdiff}
\usepackage{cleveref}
\usepackage{ifthen}

\usepackage{esdiff}

\usepackage{standalone}

\usepackage{MnSymbol} 
\usepackage{booktabs}

\usepackage{algorithm}
\usepackage{algpseudocode}

\usepackage{microtype}

\algblockdefx[Comm]{Comm}{EndComm}%
[1]{\textbf{communicate} #1}%
{\textbf{end communicate}}

\usepackage[
changebar,
todonotes,
authormarkup=none,
]{changes-pro}

\usepackage{todonotes}
\presetkeys{todonotes}{inline}{}

\hypersetup{
	pdfpagemode=UseOutlines, 
	pdfdisplaydoctitle=true, 
	pdflang=en 
}

\renewcommand{\a}{\ensuremath{_{\alpha}}}

\newcommand{\dyadic}[2]{\ensuremath{{#1} \otimes {#2}}}

\newcommand{\rz}{{\if mm {\rm I}\mkern -3mu{\rm R}\else \leavevmode
\hbox{I}\kern -.17em \hbox{R} \fi}}
\newcommand{\nz}{{\if mm {\rm I}\mkern -3mu{\rm N}\else \leavevmode
\hbox{I}\kern -.17em \hbox{N} \fi}}

\newcommand{\grad}[1]{\ensuremath{\nabla{#1}}}
\renewcommand{\div}[1]{\ensuremath{\nabla\cdot{#1}}}

\providecommand{\abs}[1]{\left\lvert#1\right\rvert}          

\newcommand{\vv}[1]{\boldsymbol{#1}}

\newcommand{\vphi}{\ensuremath{\vv{\phi}}}
\newcommand{\Jat}{\ensuremath{\vv{J}_\textup{at}}}
\newcommand{\vc}{\ensuremath{\vv{c}}}

\newcommand{\vmu}{\ensuremath{\vv\mu}}

\newcommand{\phia}{\ensuremath{\phi_\alpha}}

\newcommand{\phil}{\ensuremath{\phi_\ell}}

\newcommand{\x}{\ensuremath{\vv{\xi}}}

\newcommand{\walberla}{\textsc{waLBerla}}

\newcommand{\systemAlAgCu}{\textsf{Ag-Al-Cu}}

\def\sharedaffiliation{%
\end{tabular}
\begin{tabular}{c}}

\title{Massively Parallel Phase-Field Simulations for Ternary Eutectic Directional Solidification}

\begin{document}

\numberofauthors{1}
    \author{
      \sharedaffiliation
      {\large Martin Bauer$^1$*, Johannes H\"otzer$^{2,3}$, Philipp Steinmetz$^2$, Marcus Jainta$^{2,3}$,} \\
      {\large Marco Berghoff$^2$, Florian Schornbaum$^1$, Christian Godenschwager$^1$,}\\
      {\large Harald K\"ostler$^1$, Britta Nestler$^{2,3}$, Ulrich R\"ude$^1$} \\\\
			{\small $^1$ Chair for System Simulation, Friedrich-Alexander-Universit\"at Erlangen-N\"urnberg (FAU), Cauerstra\ss e 11, Erlangen, Germany}\\
			{\small $^2$ Institute of Materials and Processes, Karlsruhe University of Applied Sciences (HSKA),}\\ 
		        {\small Haid-und-Neu-Str. 7, 76131 Karlsruhe, Germany}\\
		        {\small $^3$ Institute for Applied Materials - Computational Materials Science (IAM-CMS), Karlsruhe Institute of Technology (KIT), } \\
		        {\small Haid-und-Neu-Str. 7, 76131 Karlsruhe, Germany}\\
		        {\tiny}\\
		        {\small *Corresponding Author. Email: martin.bauer@fau.de}\\
}

\maketitle
\begin{abstract}
Microstructures forming during ternary eutectic directional solidification processes have significant influence on the macroscopic mechanical properties of metal alloys. 
For a realistic simulation, we use the well established thermodynamically consistent phase-field method and improve it with a new grand potential formulation to couple the concentration evolution. 
This extension is very compute intensive due to a temperature dependent diffusive concentration. 
We significantly extend previous simulations that have used simpler phase-field models or were performed on smaller domain sizes. 
The new method has been implemented within the massively parallel HPC framework waLBerla that is designed to exploit current supercomputers efficiently. 
We apply various optimization techniques, including buffering techniques, explicit SIMD kernel vectorization, and communication hiding. 
Simulations utilizing up to 262,144 cores have been run on three different supercomputing architectures and weak scalability results are shown.
Additionally, a hierarchical, mesh-based data reduction strategy is developed to keep the I/O problem manageable at scale.
\end{abstract}

\section{Introduction}

The microstructure in alloys significantly influences the macroscopic properties.
In order to predict and optimize the resulting macroscopic material parameters, it is important to get a deeper insight into this microscopic structure formation.
Especially in ternary systems with three chemical species, a wide range of patterns can form in the microstructure, depending on the physical and process parameters~\cite{Ruggiero97,Lewis2002,Genau2012}. 
In this work, we show simulation results for the pattern formation in ternary eutectic directional solidification of \systemAlAgCu{} alloys.
The pattern formation of this ternary eutectic system, during directional solidification, was studied experimentally by Genau, Dennstedt, and Ratke in \cite{Genau2012,Anne2012,Dennstedt_2013}.
The thermodynamic data are reported in the Calphad database and are derived from \cite{Witusiewicz2004I,Witusiewicz2005II}.
This system is of special interest because many essential aspects of solidification processes can be studied due to the low temperature of the ternary eutectic point and similar phase fractions in micrographs.
In technical applications, this alloy is used for lead free solders in micro-electronics \cite{Kang2004}.
However, an experimental study of three dimensional structure requires significant technical effort, e.g.\ with synchrotron tomography.
Simulations of the solidification process exhibit another way of studying these structures, providing the whole microstructure evolution.

Phase-field models are well established to simulate solidification problems.
In 2011, Shimokawabe et al.~\cite{shimokawabe2011peta} presented the first peta-scale phase-field simulation on the TSUBAME 2.0 GPU-Cluster.
They studied binary dendritic growth, the evolution of multiple nuclei in three dimensions under consideration of temperature, and concentration instabilities.
Further investigations of this binary dendritic growth were performed by Yamanaka et al.~\cite{Yamanaka2011} and Takaki et al.~\cite{Takaki2013,Takaki2014} at the TSUBAME 2.0 and TSUBAME 2.5 GPU supercomputer.
In the Gordon Bell Prize 2011 awarded work, \cite{shimokawabe2011peta} used a phase-field model with two phases and two components based on a simplified free-energy approach, neglecting the temperature dependence in the concentration evolution equation and hence the slope of the solidus- and liquidus lines.
\cite{Takaki2014} continued the work of binary dendritic growth in 2D of \cite{shimokawabe2011peta} by using an improved model.
There, an anti-trapping current was added to obtain better quantitative results, but also leading to a more complex model. 
This issue of solving multiple non-linear partial differential equations and hence the increase of required memory and computational time was also mentioned by Yamanaka et al.~\cite{Yamanaka2014}.

In our work we focus on simulations of ternary eutectic directional solidfication with a phase-field model based on the grand-potential approach \cite{Diss-Choudhury201306}.
Ternary eutectic growth describes the transformation of a melt in three solid phases at a defined ternary eutectic point.
To study this ternary solidification with structures approximately two orders smaller than dendritic growth, four phases and three components have to be considered in the model.
The grand potentials, derived from thermodynamic databases, are coupled to the phase-field to ensure a physical driving force with the slopes of the solidus and liquidus planes \cite{Diss-Choudhury201306,ChoudhuryPlapp2011,Choudhury2015}.

In our simulations, we use an extension to the well established thermodynamic consistent phase-field method, including an anti-trapping current, based on a newly formulated grand potential approach \cite{hoetzer15}. 
The phase-field and chemical potential formulation consist of a relatively large number of complicated terms and parameters, making the resulting algorithm significantly more compute intensive compared to other stencil codes such as advection-diffusion calculations.
An implementation of the described method was already available in the general purpose phase-field framework \textsc{PACE3D}~\cite{Vondrous2014}, which provides a wide range of phase-field models 
coupling structural~\cite{Schneider2013,Schneider2015,Schneider2015-2}, fluid~\cite{ettrich2014fluid}, and thermal effects~\cite{Nestler2005}.
The available code was, due to its general approach, not optimized to study the considered phenomena in a most efficient way.
Simulations of large three dimensional domains are required to capture certain physically relevant effects, which can not be seen in smaller domains due to strong boundary influences \cite{hoetzer15}. 
Therefore, we re-implemented and optimized the specific model in \walberla{}, a massively parallel framework for stencil-based 
algorithms on block-structured grids~\cite{feichtinger2011walberla}. 
The framework has been shown to efficiently scale up to 1.8 million threads and run lattice Boltzmann simulations with up to one trillion cells~\cite{godenschwager2013}.

Beginning at the intra-node level by making use of SIMD instructions, to multithreading at intra-node level, up to internode parallelism via MPI, 
several layers of parallelism have to be exploited to efficiently use modern supercomputer architectures. 
We apply several optimization techniques at all these levels, resulting in an overall relative speedup of factor 80 compared to the original code
and reach approximately 25\%  of the peak FLOP rate on the SuperMUC petascale system located at LRZ Munich. 
We show scaling results on three of the largest supercomputers in Germany: on SuperMUC, JUQUEEN, and HORNET. 

The results obtained in our simulations show excellent agreement with experimental two dimensional micrographs and three dimensional tomography \cite{hoetzer15}.

\section{Model}\label{model}

To simulate the solidification process of ternary eutectic systems, a thermodynamically consistent phase-field model is used.
The presented phase-field model of Allen--Cahn type consists of two coupled evolution equations, for the vector of order parameters  $\vv \phi$ and the vector of chemical potentials $\vv \mu$.
The equations are solved using finite difference methods and an explicit Euler scheme for the time discretization. 
The domain $\Omega$ is discretized in equidistant regular cells with the spatial position ${\vv x}$.
To indicate the spatial discretization scheme of the evolution equations, we introduced the ``D$n$C$m$'' notation, as a stencil with $n$ dimensions and total access on $m$ cells.
%
%
In the domain $\Omega$, the order parameters  $\phi_\alpha ({\vv x},{t})$, $\alpha \in [1,N]$, describe the fractions of the $N$ thermodynamic phases in each cell and each time step ${t}$.
We define $\vv\phi({\vv x},{t}) $ on the regular $N-1$ simplex $\Delta^{N-1} , \forall {\vv x}, {t}$.
The evolution equations of the phase-fields are written as
\begin{equation}
\diffp{\phia}{t} = \frac{1}{\tau_\alpha\epsilon} ( \text{rhs}_\alpha - \frac{1}{N} \sum_{\beta=1}^N \text{rhs}_\beta )
 \label{func:phi}
\end{equation}
with
\begin{align}
\text{rhs}_\alpha &= 
\underbrace{\epsilon T \left(\diffp{a(\vv\phi,\nabla\vv\phi)}{\phia} - \nabla \cdot \frac{\partial a(\vv\phi,\nabla\vv\phi)}{\partial\nabla\phia} \right)}_\text{D3C7} 
\nonumber\\
&+ 
\underbrace{\frac{1}{\epsilon} T \diffp{\omega(\vv\phi)}{\phia}}_\text{D3C1}
+ \underbrace{\diffp{\psi(\vv\phi,\vv\mu,T)}{\phia}}_\text{D3C1} 
.
\label{func:phi_rhs}
\end{align}
The reciprocal relaxation parameter $\tau_\alpha$ couples the phase-field evolution with the physical time scale and $\varepsilon$ is related  to  the interface width.
The gradient energy density $a(\vv\phi,\nabla\vv\phi)$ and the potential energy density $\omega(\vv\phi)$ describe the interfacial energy part in \eqref{func:phi_rhs}.
The driving force $\psi(\vv\phi,\vv\mu,T)$ describes the thermodynamic process of the phase transition caused by the undercooling and connects the evolution of the order parameter to the chemical potential.
This force is derived by parabolically fitted Gibbs energies which are derived from the thermodynamic Calphad databases \cite{Choudhury2015}.
The evolution equations of the $K$ chemical potentials $\vv \mu({\vv x},{t})$ are derived from Fick's law and the variational derivation of the concentrations $\vv c({\vv x},{t})$.
Due to the conservation of mass, one concentration can be derived from the others. 
The number of evolution equations $\vv c({\vv x},{t})$ as well as the number of components of $\vv \mu({\vv x},{t})$ can therefore be reduced to $K-1$.

The $\vv \mu$ evolution is written as
\begin{align}
\diffp{\vmu}{t}  =& \underbrace{\left[ \left( \diffp{\vc}{\vmu} \right)_{T,\vphi}\right]^{-1} }_\text{D3C1}  
       \bigg( 
        \underbrace{ - \left( \diffp{\vc}{\vphi} \right)_{T,\vmu} \diffp{\vphi}{t} - \left( \diffp{\vc}{T} \right)_{\vmu,\vphi} \diffp{T}{t}  }_\text{D3C1}  \nonumber \\  
        & +\underbrace{ \div{\left(\vv{M}(\vv \phi,T)\grad{\vmu} \right)} }_\text{D3C7} 
- \underbrace{ \div{ \left(\Jat(\vphi,\vmu,T) \right)}  }_\text{D3C19}
       \bigg)
\label{func:mu}
\end{align}
The Jacobian matrix $\partial{\vc}/\partial{\vmu}$ describes the susceptibility and is derived from parabolic free energies.
The two flux terms are a gradient flux of the chemical potential depending on the mobility $\vv M(\vv \phi, T)$ and a flux, called \emph{anti-trapping current} \cite{Diss-Choudhury201306}.
It is derived as
\begin{align}
 \vv{J}_{at} = \frac{\pi \epsilon}{4} \sum_{\substack{\alpha=1 \\ (\alpha\neq \ell)}}^N
             \frac{g\a(\vphi)h_{\ell}(\vphi)}{\sqrt{\phia\phi_{\ell}}} \diffp{\phia}{t}
             \left(\frac{\grad \phia}{\abs{\grad \phia }} \cdot \frac{\grad \phil}{\abs{\grad \phil }} \right) 
            \nonumber\\              
             \left(\dyadic{\left(\vc^\ell(\vmu)-\vc^\alpha(\vmu)\right)}{\frac{\grad \phia}{\abs{\grad \phia }}} \right)
\label{eq:antitrap}
\end{align}
with the interpolation function $h_{\ell}$ \cite{moelans2011}.

For the directional solidification, we use a frozen temperature assumption by imprinting an analytical temperature gradient with a defined velocity. 

For the description of the numerical optimizations, we distinguish different regions in the simulation domain.
The region $B_\alpha := \{\vv x \in \Omega \,|\, \phia(\vv x, t) = 1 \wedge |\vv\nabla \phia| = 0 \}$, where only one phase $\alpha$ exists, is called \emph{bulk region}.
In this region, the time derivative of the order parameter and the anti-trapping current \eqref{eq:antitrap} are zero.
 
The diffuse interface $I_\Omega := \Omega \setminus  \bigcupdot_{\alpha \in N} B_\alpha$ is located between bulk regions.
The interfacial energy part and the driving force in \eqref{func:phi} are exclusively calculated in $I_\Omega$. 
The solidification front is defined as $F_\Omega := \{\vv x \in I_\Omega \; |\; \phi_\ell(\vv x) > 0\}$.
The liquid region is defined as $L_\Omega := B_\ell$ and the solid one as $S_\Omega := \Omega \setminus ( L_\Omega \cup F_\Omega)$.
For a more detailed description of the model we refer to \cite{hoetzer15,Diss-Choudhury201306}.

\subsection{Phase-field Algorithm}
\begin{figure}%
  \centering
  \includegraphics[width=0.9\columnwidth]{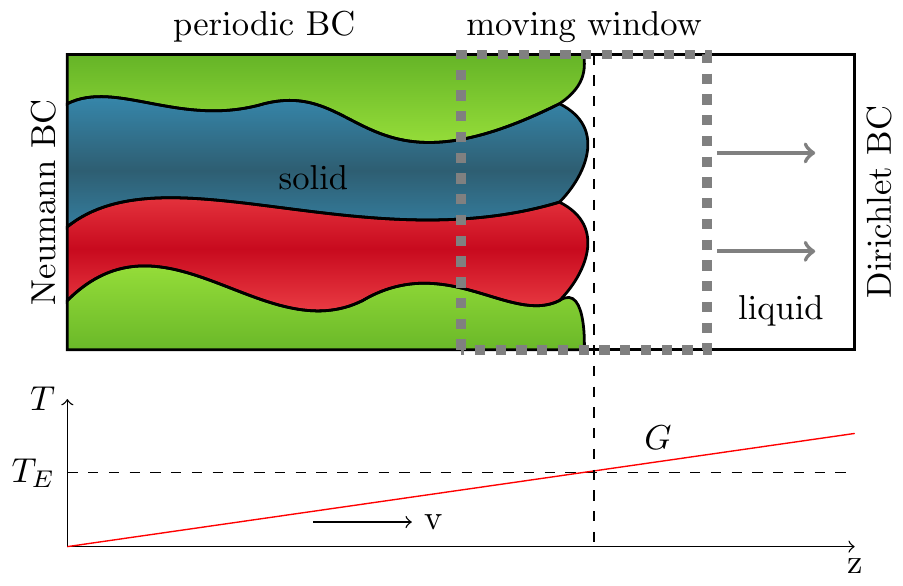}
  \caption{Simulation setting for the directional solidification of a ternary eutectic system including the boundary conditions. Below, the analytical temperature gradient $G$ with the velocity $v$ and its direction is shown.}
  \label{fig:sim-setting}
\end{figure}

The algorithm to calculate the phase-field model is decomposed into two kernels: one for updating values of the phase-field itself \eqref{func:phi} and one to update the chemical potential \eqref{func:mu}. 
Two lattices are allocated for each variable: two destination fields denoted by $\phi_\text{dst}$ and $\mu_\text{dst}$ and two source fields, denoted by $\phi_\text{src}$ and $\mu_\text{src}$.
In the destination fields values for the next time step $t+\Delta t$ are stored, while source fields hold values for the current time step.
Each kernel performs a loop over the local simulation domain and updates all cell values of the $\phi$ or $\mu$ field.
After each kernel run, the ghost layers are synchronized with neighboring blocks and boundaries are updated using Dirichlet, Neumann and periodic boundary conditions as shown in Figure~\ref{fig:sim-setting}.
The details of this update scheme are depicted in Algorithm~\ref{alg:phasefield}.

\begin{algorithm}
  \caption{Timestep}
  \label{alg:phasefield}
      \begin{algorithmic}[1]
      
      \State $\phi_\text{dst} \gets \mathbf{\phi}\mbox{-kernel}  \Big( \phi_\text{src}, \mu_\text{src}     \Big) $ (see \eqref{func:phi})

      \State $\phi_\text{dst}$ ghostlayer communication
      \State $\phi_\text{dst}$ boundary handling
      
      \State $\mu_\text{dst}  \gets \mathbf{\mu}\mbox{-kernel} \Big( \mu_\text{src}, \phi_\text{src}, \phi_\text{dst}    \Big) $ (see \eqref{func:mu})
      
      \State $\mu_\text{dst}$ ghostlayer communication
      \State $\mu_\text{dst}$ boundary handling
      \State Swap $\phi_\text{src} \leftrightarrow \phi_\text{dst}$  and $\mu_\text{src} \leftrightarrow \mu_\text{dst}$
      \end{algorithmic}
\end{algorithm}

\begin{figure}[htp]
  \subfigure[$\phi$-sweep]{\includegraphics[width=0.45\columnwidth]{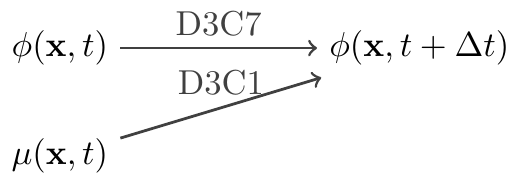}}
  \hspace{0.099\columnwidth}
 \subfigure[$\mu$-sweep]{\includegraphics[width=0.45\columnwidth]{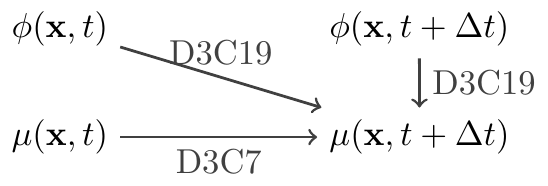}}
 \caption{Data dependencies of the two kernels.}
 \label{fig:data_dependencies_no_overlap}
\end{figure}

The $\phi$-kernel needs the previous phase $\phi_\text{src}$ and chemical potential $\mu_\text{src}$ values as input. 
Direct neighborhood values are required to compute gradients of the phase-field, leading to a D3C7 stencil for the $\phi$-field while only local values are required for $\mu$ (D3C1).
For updating the chemical potential, previous chemical potential values $\mu_\text{src}$ as well as the phase-field values of two time steps ($\phi_\text{src}$ and $\phi_\text{dst}$) are necessary.
The two phase-field timesteps and the D3C19 stencil, including diagonal neighbors, are required to calculate \eqref{eq:antitrap}.
Figure~\ref{fig:data_dependencies_no_overlap} shows these data dependencies in detail for the two kernels.



As initial setup we use solid nuclei at the bottom of a liquid filled domain as shown in Figure~\ref{fig:sim-setting}.
These solid nuclei are created by a Voronoi tesselation with respect to the given volume fractions of the phases.
The simulation parameters of \cite{hoetzer15} are used for our studies in the presented model, describing the system \systemAlAgCu{}.

\section{Implementation}
In the following section, we present our implementation of the phase-field model and apply 
optimizations necessary to efficiently run on current supercomputing systems.

\subsection{The waLBerla Framework}

The phase-field method described above was implemented in the \walberla{} framework (widely applicable lattice Boltzmann from Erlangen). 
As the name suggests, the framework was initially developed for simulations using the lattice Boltzmann method.
Over time, it evolved into a multi-physics framework for efficient implementations of stencil-based algorithms on block structured grids in a massively parallel way.
\walberla{}  partitions the simulation domain into equally sized chunks, called blocks. 
On each block, a regular grid is allocated, extended by one or more ghost layers for communication in MPI parallel simulations.
This concept provides the flexibility for supporting complex geometries \cite{godenschwager2013} and optional mesh refinement.
The regular structure inside the blocks allows for highly efficient compute kernels.
The data structure storing the blocks is fully distributed: Every process holds information only about local and adjacent blocks. Thus, the memory usage of
a process does not depend on the total size of the simulation but only on the size of local and neighboring blocks.
Only during the startup phase which is responsible for block setup and load balancing, one process has to store global domain information. 
This initialization can be executed independently of the actual simulation. The resulting block structure is then stored in a file to be
loaded by the simulation at runtime.
The framework is entirely written in C++ with small portions of the code being automatically generated.
For performance critical code sections, we make use of template meta programming techniques to achieve good performance without losing
flexibility and usability. 


\subsection{Input/Output}

A big challenge in massively parallel programs is dealing with the huge amount of output data.
In many big simulations, I/O operations can become a significant bottleneck.
Therefore, we try to minimize the amount of data that is read and written to the file system.
Since the initial domain setup is generated using a Voronoi approach, we do not have to load big, voxel-based input files
describing the domain. I/O is only necessary for checkpointing and for result output.
For generating checkpoints, the complete simulation state has to be stored on disk, containing four $\phi$ values and
two $\mu$ values per cell. While all computations are carried out in double precision, checkpoints
use only single precision to save disk space and I/O bandwidth. 
Writing a checkpoint can take a significant amount of time 
compared to a simulation time step, therefore checkpoints are written infrequently.

Simulation results have to be written more often, thus a faster technique is required for this task.
Instead of writing all values of a cell, we only store the position of the interfaces using a triangle surface mesh.
Therefore, a custom marching cubes algorithm based on \cite{lorensen1987marching} was developed that generates meshes locally on each block, using the $\phi$ values as input.
For each phase, a mesh is generated, describing the interface between this phase and all other phases.
The marching cubes algorithm extends to the ghost regions such that the local meshes can be stitched together to a single mesh 
describing the complete domain. These local meshes could be written to disk, leaving the stitching for postprocessing.
Since the marching cubes algorithm creates triangles with edge lengths in the order of $dx$, these meshes are still unnecessarily fine
and could be adaptively coarsened without losing much accuracy. This coarsening step can optionally be performed before writing the mesh 
to keep output file sizes small.
For mesh coarsening, we use the quadric-error edge-collapse-based simplification algorithm \cite{garland1997surface} from the
\emph{Visualization and Computer Graphics Library} (VCG) \cite{vcglibrary}. The coarsening is done in a hierarchical way:
In a first step, each process calls the edge-collapse algorithm on its local mesh. 
By assigning a high weight to all vertices that are located on block boundaries, the boundaries are preserved
such that the later stitching step can work correctly. 
Then, two local meshes are gathered on a process, stitched together, and again coarsened in the stitched region. 
This step is repeated $\log_2(\textup{processes})$ times where in each step only half of the processes take part in the reduction.
This procedure is stopped if the mesh is either fully reduced or has reached a size that cannot be stored in the memory of a single node.
In the latter case, postprocessing can be resumed on a machine with more memory than a compute node.

\subsection{Optimizations}
\label{sec:Optimizations}
Starting with a very general phase-field model, the goal of this work was to develop a highly optimized code for the specific model, presented in \cite{Diss-Choudhury201306}, which efficiently can simulate sufficiently large 3D domains. 
Optimizations are done on different levels by exploiting physical, mathematical, and computational facts to decrease the computation time.

For directional solidification, we can reduce the effective domain size in the solidification direction ($Nz$) \cite{hoetzer15} by using a moving window technique \cite{Vondrous2014} as shown in Figure~\ref{fig:sim-setting}.
The evolution in the solid is multiple magnitudes lower than in the liquid such that we neglect the evolution in the solid. 
In order to produce physically correct microstructures, the base size of the domain $Nx \times Ny$ requires a minimal size to reduce the influence of the periodic boundaries. 

The evolution of the temperature can be described by a frozen temperature ansatz in solidification direction, by exploiting the time scales of the different evolution equations \cite{hoetzer15,Diss-Choudhury201306}. 
At a given time $t$, we assume the temperature being constant in slices orthogonal to the solidification direction.

The grand potentials for the driving force are only needed in the range of the ternary eutectic point, therefore we use fitted parabolic Gibbs energies to derive the potentials instead of describing the total ternary system. 
This simplifies the calculations involving the concentrations $\vv c$ and the chemical potentials $\vv \mu$~\cite{Choudhury2015}. 

The model equations can be simplified for certain parts of the domain for the two evolution equations. 
The computation of the anti-trapping current in the $\mu$-evolution is only required in the interface region $F_\Omega$ of the solidification front.
Since evaluating this current is computationally expensive, we detect as early as possible if the complete term has to be evaluated by testing critical subexpressions for zeros.
When computing $\Jat$, we first check if $\vphi$ is zero since then $h_\ell(\vphi)$ and thus $\Jat$ are zero.
By introducing one additional check, we can omit the calculation of the expensive $\Jat$ in all cells which do not contain liquid.
A similar check can be added for $\grad \vphi_l$: for cells with zero liquid phase gradient, the computation of the anti-trapping current can be skipped as well.
The interface region $I_\Omega$ is bounded due to a sinus-shaped interface profile in which $\nabla \phi_\alpha \neq 0$. 
Following the calculation of the $\phi$ evolution is only required in this small interface region.
Adding these kind of checks introduces possibly expensive \texttt{if} conditions in the kernel, so there may be a trade-off between the saved computations and the peak performance.
As shown in section \ref{sec:results}, the introduced checks reduce the total runtime considerably.

We conducted further optimizations on the source code and hardware level using processor extensions available on the target machines.
On these levels, we used a systematic, performance model driven approach to identify and eliminate relevant bottlenecks.
As we show in section \ref{sec:results}, our code is bound by in-core execution, therefore the goal is to reduce the total number of floating point operations, potentially at the cost of increased memory traffic.
 
The fact that in our specific scenario the temperature is a function of the $z$ coordinate led us to choose the $z$ iteration as the outermost loop in  computation kernels, followed by loops over $y$ and $x$.
Since many values, which are required to calculate the driving force $\psi$ of \eqref{func:phi_rhs} or the anti-trapping current $\Jat$ of \eqref{eq:antitrap}, depend on analytic temperatures only, these values can be pre-calculated once for each $z$-slice instead of computing them again in every cell.

Following the strategy of saving floating point operations at the cost of memory transfers, we identify expressions in the model equations that are evaluated multiple times and can therefore be buffered.
In our discretization scheme we also evaluate quantities at staggered grid positions. 
To avoid evaluation in each cell, they are written to a buffer after computation and are reused.
Since the outermost loop in the computation kernels iterates over the $z$ coordinate, a buffer of the size $Nx\times Ny$ is needed.

\begin{figure}%
  \centering
   \includegraphics[width=0.9\columnwidth]{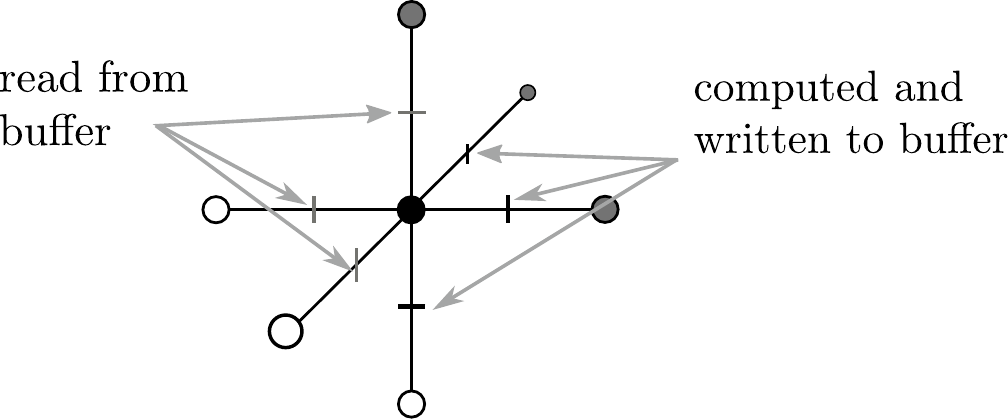}
  \caption{Buffering values at staggered positions (D3C7). Circles mark cell midpoints, lines indicate cell boundaries. }
  \label{fig:staggeredBuffer}
\end{figure}

The computationally most intensive part of equation (\ref{func:mu}) is the calculation of the divergence of $\vv{v}_\textup{buf}:=(\vv{M}\grad{\vmu}-\Jat)$, even if the evaluation of the anti-trapping current can be skipped for many cells.
In order to update one cell, the six values of $\vv{v}_\textup{buf}$ at neighboring staggered positions are required.
However, three of them can be buffered and reused since they have already been calculated during the update of previous cells (Figure~\ref{fig:staggeredBuffer}).
Only in cells located at block boundaries, all six values have to be computed explicitly.
Using this buffering approach, we effectively halve the required floating point operations at the cost of additional memory accesses for reading and writing the buffered values.

The same technique of buffering staggered values is also applied for evaluating the divergence of $\partial a(\vv\phi,\nabla\vv\phi)/\partial\nabla\phia$ in the phase-field update step.

While adhering to the established data parallel approach on inter-process level, the \walberla{} framework offers task parallel programming models on intra-process level to support overlapping communication with computation.
The computation kernels as well as the ghost layer exchange routines are implemented as C++ functors, which are registered at a ``Timeloop'' class to manage the communication hiding.

The communication of the chemical potential field can be hidden in a straightforward way since the
following update of the phase-field only depends on local $\mu$ values (Figure~\ref{fig:data_dependencies_no_overlap}). 
This allows us to communicate the $\mu$ boundaries while updating the $\phi$-field. 
In our implementation the order of communication and boundary handling routines can also be interchanged without altering the results. 

\begin{figure}
  \centering
  \includegraphics[width=0.9\columnwidth]{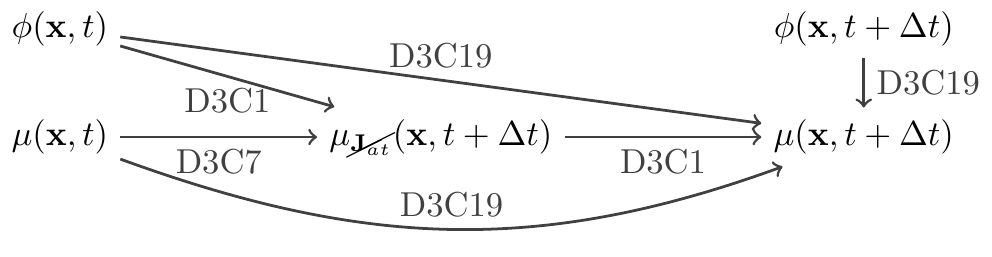}
  \caption{Data dependencies (with communication overlap)}
  \label{fig:data_dependencies_overlap}
\end{figure}

Since updating $\mu$ accesses neighborhood values of $\phi$ for the current and new time step, the communication of the phase-field values can not simply be overlapped with $\mu$ computation. 
In order to hide the phase-field communication as well, the $\mu$ update step has to be split up into two kernels. 
The obvious method would be to update inner values first and after communication has finished, update the values at the border.
Implementing this solution would require a staggered value buffer of the same size as the complete field or recomputation of staggered values. 
Thus, we split up the $\mu$ update into a local and a non-local $\phi$ dependency as indicated in \eqref{func:mu}. 
This splitting simplifies the data dependencies (Figure~\ref{fig:data_dependencies_overlap}) and does not affect the buffering of staggered values, but still has some overhead, since now the temperature dependent values have to be computed twice for each $z$-slice.
Therefore, we calculate the $\mu$ evolution without the anti-trapping current \eqref{eq:antitrap} first. 
After transfer of the ghost layers, the anti-trapping current is calculated and added to the  $\mu$ evolution.
Algorithm \ref{alg:phasefield_overlap} shows the resulting time step of the phase-field method 
using fully overlapping communication.

\begin{algorithm}
  \label{alg:phasefield_overlap}

  \caption{Timestep with communication overlap}

      \begin{algorithmic}[1]
      \Comm{ $\mu_{src}$ }
	\State $\phi_\text{dst} \gets \mathbf{\phi}\mbox{-sweep}     \Big(     \phi_\text{src} , \mu_\text{src}  \Big) $
      \EndComm
            \vspace{0.2cm}

      \State $\phi_\text{dst}$ boundary handling
      
      \Comm{ $\phi_\text{dst}$ }
      \State $\mu_\text{dst}  \gets \mathbf{\mu}\mbox{-sweep-local}   \Big(     \mu_\text{src} , \phi_\text{src}, \phi_\text{dst}    \Big) $
      \EndComm
      \vspace{0.2cm}

      \State $\mu_\text{dst}  \gets \mathbf{\mu}\mbox{-sweep-neighbor}   \Big(    \mu_\text{src} , \phi_\text{src}, \phi_\text{dst}    \Big) $
     
      \State $\mu_\text{dst}$ boundary handling
      \State Swap $\phi_\text{src} \leftrightarrow \phi_\text{dst}$  and $\mu_\text{src} \leftrightarrow \mu_\text{dst}$
      \end{algorithmic}
\end{algorithm}

In order to fully utilize modern hardware architectures, especially when running compute intensive codes, it is important to make full use of their SIMD (single instruction, multiple data) capabilities.
While most compilers can do automatic loop vectorization, it is still necessary to provide additional information to the compiler regarding aliasing, data alignment, and typical loop sizes using non portable pragma directives.
An analysis with the performance measurement tool \emph{LIKWID} \cite{likwid}, which can measure the amount of vectorized and non-vectorized floating point operations, showed that large portions of our code were not automatically
vectorized. While the pragma approach proves useful for small concise loop kernels, we could not improve the more complex compute kernels for $\phi$ and $\mu$ of several hundred lines of code using this approach. 
Instead, we explicitly vectorize these two main compute kernels with intrinsic compiler functions. Using intrinsics requires a manual reformulation of the algorithm in terms of vector data types. 
Since these intrinsic functions directly map to assembler instructions, they are not portable across different architectures. 
In order to keep the code portable, a lightweight abstraction layer was developed that provides a common API supporting the Intel processor extensions SSE2, SSE4, AVX, AVX2 and the QPX extension on Blue Gene/Q cores.
Not all functions of this API directly map to a single intrinsic function and therefore to a single assembler instruction for each instruction set. 
For example, AVX2 provides instructions for permuting elements of vector data types, which require two or more instructions in the SSE extension. 
Our API hides these differences, providing functions for all AVX2 and QPX intrinsics which are emulated on older extensions
in the most efficient way. Manual inspection of the assembler code shows that calls to this thin SIMD API are inlined by the compiler and therefore no additional overhead is introduced.


\section{HPC Systems}

In this section we shortly describe the three Tier-0/1 cluster systems at the High Performance Computing Center Stuttgart (HLRS), the Leibniz Supercomputing Center (LRZ), and the Jülich Supercomputing Centre (JSC).
All systems support vectorization, either using AVX(2) on the Intel based chips or QPX on BlueGene/Q cores.

\subsection{SuperMUC} 

The SuperMUC system located at Leibnitz Supercomputing Center in Munich is built out of 18,432 Intel Xeon \mbox{E5-2680} processors running at 2.7\,GHz resulting in a total number of 147,456 cores \cite{supermuc}.
One compute node of SuperMUC consists of 2 sockets, each equipped with 8 cores.
512 nodes are divided into one island. Within each island, SuperMUC uses a non-blocking tree network topology, whereas all 18 islands are connected via a pruned tree (4:1) \cite{supermuc}.
On SuperMUC we used the Intel compiler in version 14.0.3 together with IBM MPI.
We compiled our code on optimization level 3 together with optimization flags enabling interprocedural optimization at link time and fast floating point operations.

\subsection{Cray XC40 (Hornet)}  
The Cray XC40 system at the HLRS consists of 3944 nodes with two sockets, each containing a Intel Haswell E5-2680v3 with 12 cores. 
The systems contains 94,656 cores in total which are interconnected by a Dragonfly network\cite{dragonfly} called  Cray Aries. 
The Haswell core clock rate is $2.5$ GHz and it supports Hyper-Threading as well as the newer AVX2 instruction set. 
On Hornet we have chosen the same compiler and compiler options as on SuperMUC with additional flags to make use of the AVX2 extension. 

\subsection{JUQUEEN }

JUQUEEN is, as of April 2015, the fastest supercomputing system in Germany and is positioned on rank 8 on the TOP500 list\cite{top500}.
It is a 28 rack Blue~Gene/Q system providing 458,752 PowerPC A2 processor cores, with each core capable of \mbox{4-way} multithreading \cite{gilge2014ibm}.
JUQUEEN uses a 5-dimensional torus network topology capable of achieving up to 40\,GB/s, with latencies in the range of a few hundred nanoseconds.
We compiled our code with the IBM XL compiler in version 12.1 with the compiler optimization level set to 5.

\section{Results and Discussion}
\label{sec:results}

In this section, we present single core performance results as well as scaling experiments run on these three supercomputing systems.
The presented performance results are measured in MLUP/s, which stands for ``million lattice cell updates per second''.
At the end of this section, we show simulation results of the \systemAlAgCu{} system and compare them to experimental data.

\subsection{Performance Results}

Since the performance of the compute kernels depends on the composition of the simulation domain as shown in section \ref{sec:Optimizations}, we executed our benchmarks separately for three representative parts of the domain.
These different scenarios are labeled \emph{interface}, \emph{solid}, and \emph{liquid} and correspond to the regions $F_\Omega$, $L_\Omega$, and $S_\Omega$ defined in Sec.~\ref{model}. 
The \emph{solid} scenario consists purely of already solidified material, as is the case in a realistic simulation in the lower third of the domain.
The \emph{interface} scenario simulates only the solidification front, i.e.\ the middle third of the simulation domain.
The upper part of the domain consists only of liquid phase and is represented by the \emph{liquid} scenario.


\subsubsection{Single Core Performance}

The starting point of our HPC implementation was a general phase-field code written in C. 
One main design goal of this application code is flexibility to allow rapid prototyping and testing of new models. 
A wide range of models is already implemented in this code: various phase-field models, a structural mechanics as well as a fluid mechanics solver, which can all be coupled with each other.
As already mentioned above, for the specific model described here, simulations of large three dimensional domains are necessary to obtain physically meaningful results. 
This motivates the design and implementation of a new code that is highly optimized and parallelized to exploit the largest HPC systems available.

\begin{figure*}%
  \centering
   \includegraphics[width=0.49\textwidth]{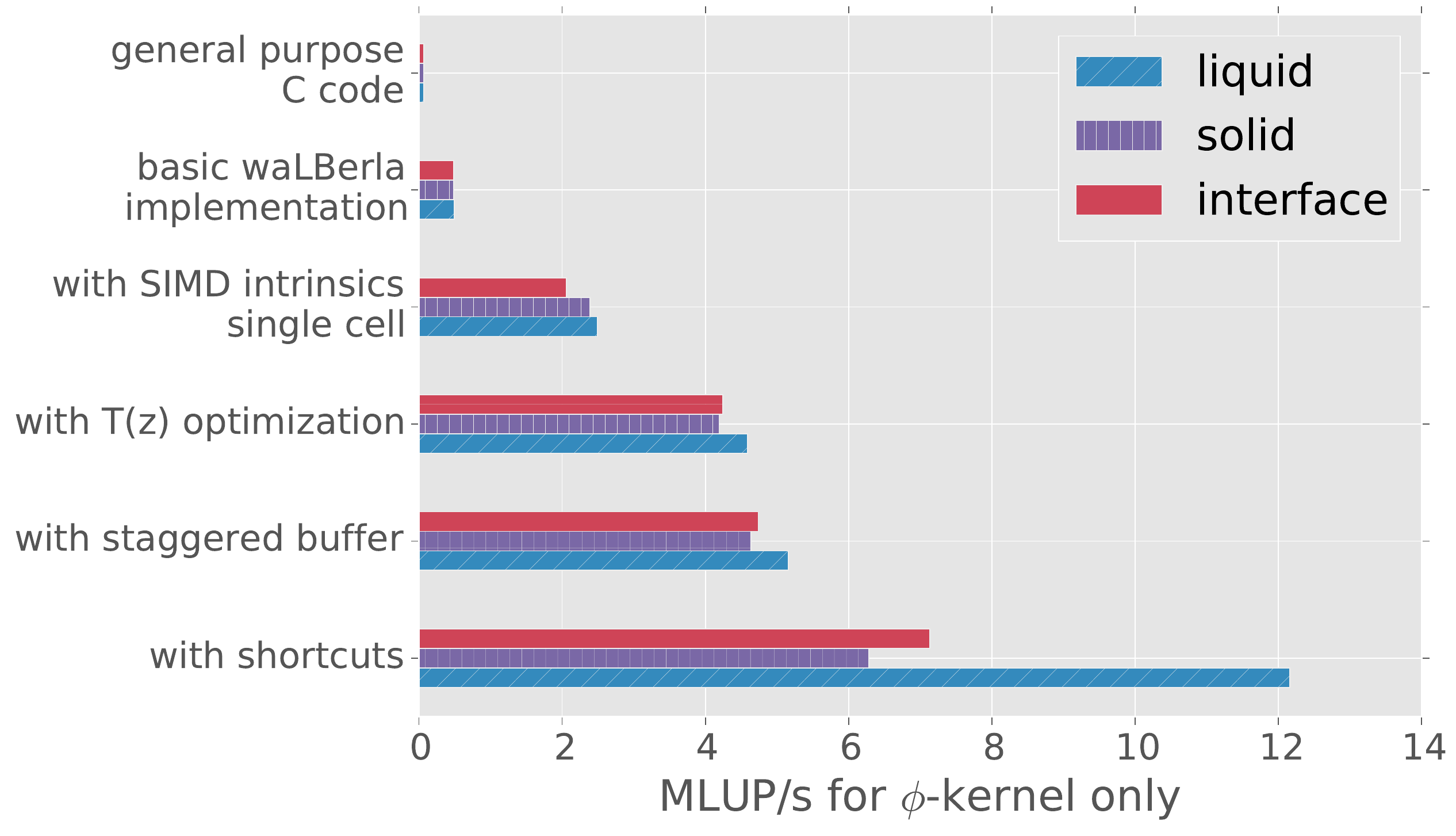}
   \includegraphics[width=0.49\textwidth]{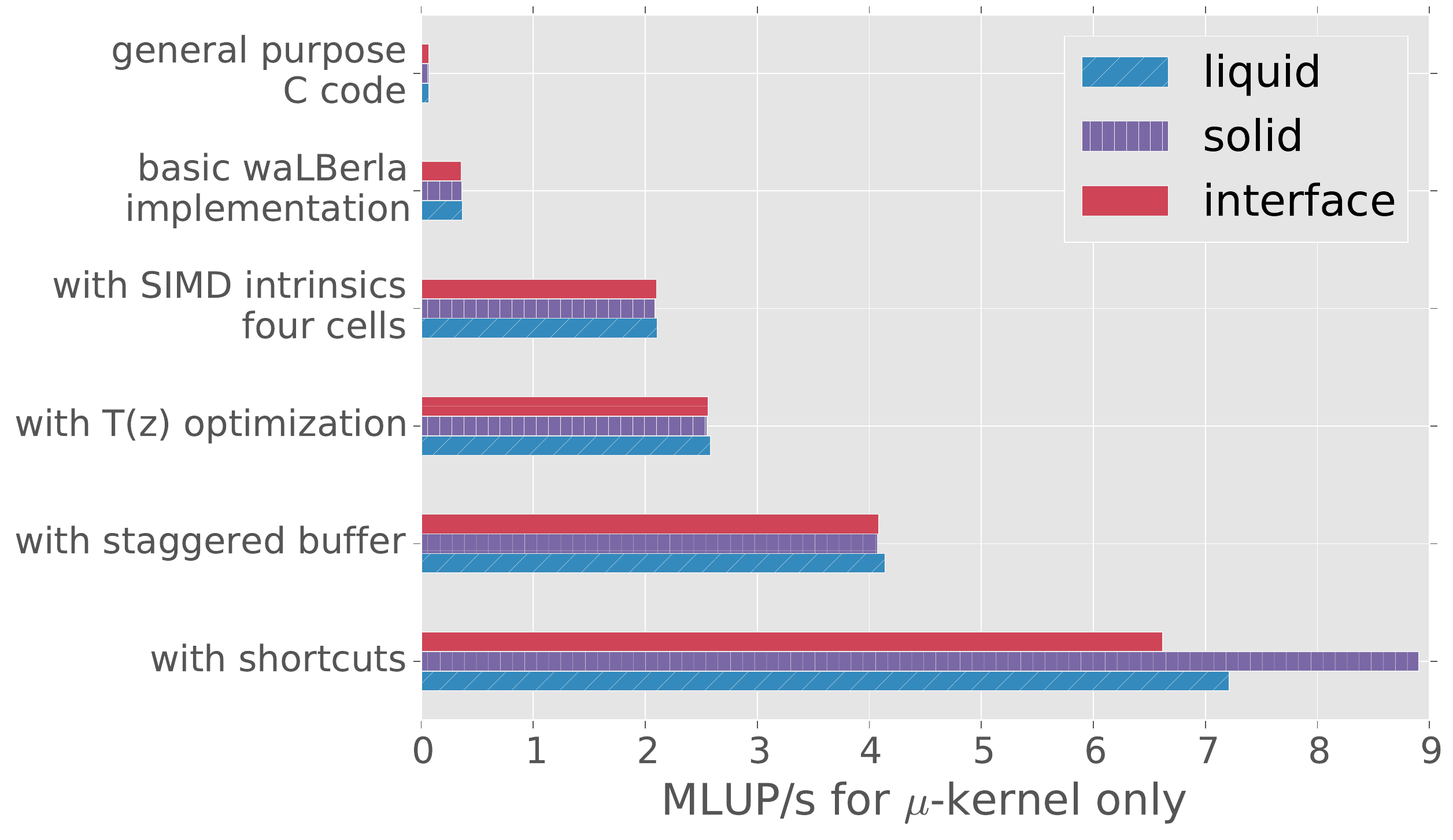}  

  \caption{ Optimization for $\phi$-kernel (left) and $\mu$-kernel (right), run in {\em interface} blocks, block size $60^3$ }
  \label{fig:sweep-optimizations}
\end{figure*}

A straightforward re-implementation of the grand chemical potential based phase-field model in the \walberla{} framework already yields significant performance improvements (Figure~\ref{fig:sweep-optimizations}).
Since this new implementation is targeted at one specific model, certain basic optimizations can already be applied easily to the compute kernels: While the original implementation makes heavy use of indirect function calls via function pointers at cell level, we either remove these indirections in the \walberla{} version entirely due to the specialization of the code to one specific model, or replace them with static polymorphism using C++ templates. 
Additionally, we replaced divisions where the denominator is guaranteed to be in a small set of values by table lookup and multiplication with the inverse. The number of division operations is further reduced by replacing inverse square root calculations, required for vector normalizations, with approximated values provided by a fast inverse square root algorithm~\cite{lomont2003fast}.

In a second step we implement more advanced, aggressive optimizations. These include explicit SIMD vectorization in both computation kernels.
Since the targeted architectures all have a vector width of four double precision values, the straightforward way for vectorizing the algorithm is to unroll the innermost loop, updating four cells in one iteration.
While this technique is the only possible one for the $\mu$-kernel, a more natural approach exists for the $\phi$-kernel of our specific model: Instead of handling four cells simultaneously, we can use a SIMD vector to represent the four phases of a cell. With this technique, the field is still updated cellwise, such that branching on a cell-by-cell basis becomes possible. This branching can significantly speed up the kernels, since some expensive terms have to computed only for certain cell configurations. Vectorized kernels that handle four cells simultaneously can only take these shortcuts if the condition is true for all four cells. Since the single cell kernel version operates on less data per iteration compared to the four cell version, more intermediate values can be kept in vector registers, which would have to be spilled to memory otherwise.
A drawback of the {\em cellwise} version is the need for various permute or rotate operations when computing terms that contain single components of the $\phi$ vector. This happens for example when expressions like $\phi_\alpha \sum_{\beta=1}^4 \phi_\beta $ have to be evaluated. 

\begin{figure}
  \centering
   \includegraphics[width=\columnwidth]{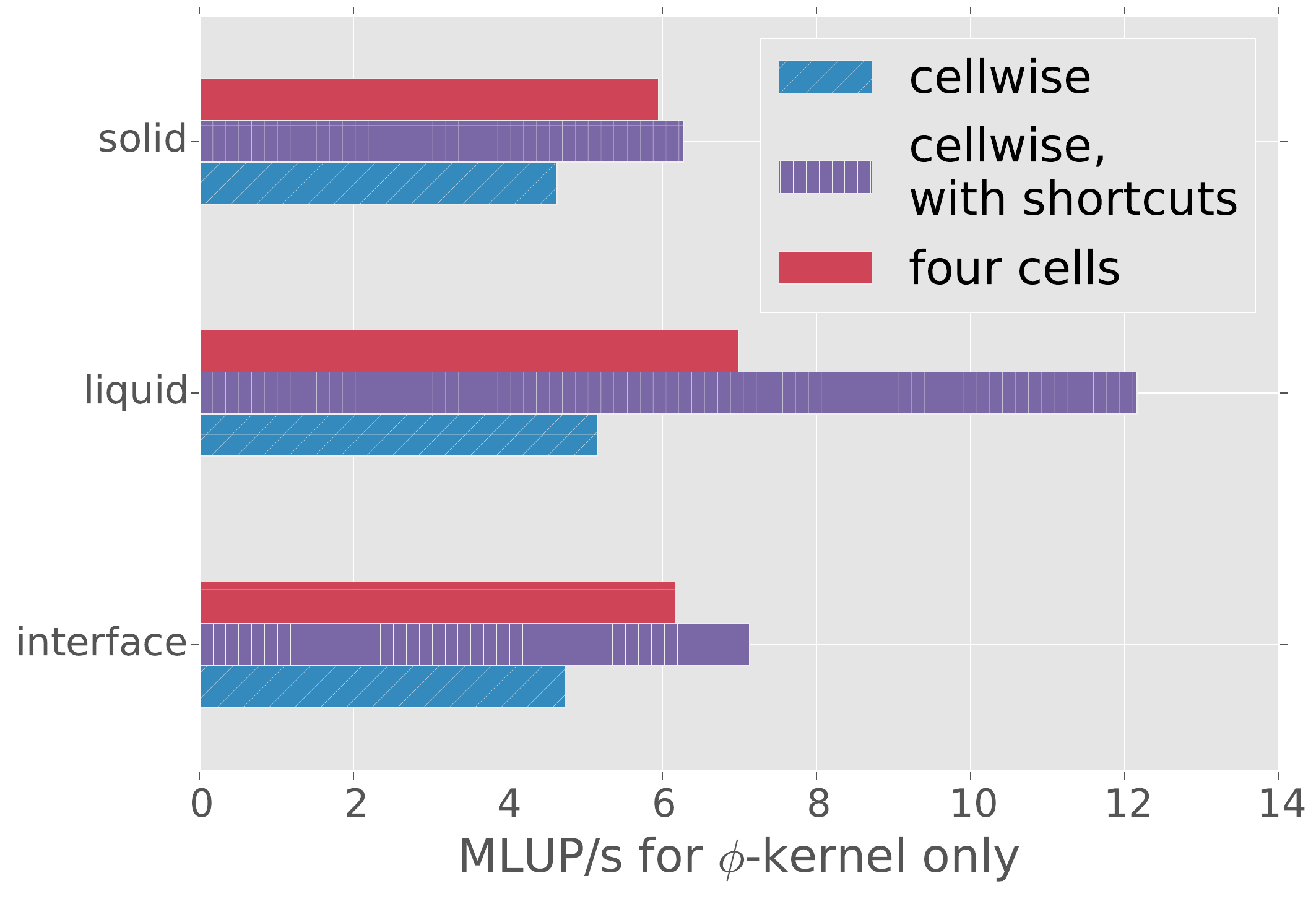}
  \caption{Comparison of different vectorization strategies on one SuperMUC core, block size chosen as $60^3$.}
  \label{fig:simd_comparison}
\end{figure}

Figure~\ref{fig:simd_comparison} shows the performance of three vectorized $\phi$-kernel implementations: The first two implementations both use the cellwise vectorization approach. While the first version evaluates all terms for each cell, the second version contains tests that determine which terms have to be evaluated ({\em cellwise with shortcuts} ). The third version uses {\em four-cell-vectorization} and can only skip terms that are not required for all of the four cells. 
In all three parts of the domain, the single cell kernel with shortcuts performes best. The additional costs of the vector permute operations are alleviated by the ability to branch on a per cell basis and the reduced number of required SIMD registers. 
For all further benchmarks, we use this fastest cellwise kernel with shortcuts to update the phase field.
Since we apply two different vectorization strategies to the two compute kernels, the optimal data layout of the $\phi$-field for the two kernels is different: 
The $\mu$-kernel processes the data four cells at a time, so a structure-of-arrays (SoA) layout would be the best choice, while the fastest $\phi$-kernel requires an array-of-structures (AoS) layout 
to be able to load a SIMD vector directly from contiguous memory. We choose a SoA layout since the $\mu$-kernel has to load phase field values of 38 cells (19 cells from $\phi(x,t)$ and 19 from $\phi(x,t+\Delta t)$) whereas the
$\phi$-kernel only has to load 7 cells. Due to the high computational intensity of the $\phi$-kernel, no notable  differences could be measured in the $\phi$-kernel performance after a data layout change of the $\phi$-field.
Figure~\ref{fig:sweep-optimizations} shows the increase in performance of the new SIMD kernels compared to the basic implementation.
By applying vectorization only, a maximal speedup of factor 4 can be expected. The high speedup of factor 5 to 7 is due to the fact that in addition to vectorization also additional optimizations like common subexpression precomputation were done at this stage.
This comes at the cost of decreased flexibility: While in the basic implementation the code was structured according to the mathematical formulation of the model, the single terms of the model can hardly be recognized in the manually vectorized SIMD kernels.
To decrease the maintenance effort for the various kernels, a regularly running test suite checks all kernel versions for equivalence.

Figure~\ref{fig:sweep-optimizations} additionally shows the performance improvement due to optimizations we can make
when the temperature is prescribed as a function of time and one space component $z$ only. In this case, we precompute all 
temperature dependent terms once for each x-y-slice instead of computing them in each cell.
Using this optimization increases the performance of the  $\mu$-kernel by approximately 20\%  
and the performance of the $\phi$-kernel by 80\%.

While this technique improved predominantly the runtime of the $\phi$-kernel, the staggered buffer optimization is targeted 
at the $\mu$-kernel and the costly computation of the term $\div{\left(\vv{M} \grad{\vmu} - \vv{J}_{at} \right)} $ which is evaluated 
at staggered positions in the grid. Buffering and reusing half of these values increases the $\mu$-kernel performance by almost a 
factor of two, which shows that the runtime of this kernel is dominated by the calculation of these staggered values.
In the $\phi$-kernel, the same technique is applied for buffering gradient values at staggered positions. In this case, the 
buffered values are not as expensive to compute as the buffered quantities in the $\mu$-kernel, therefore this optimization leads 
only to slightly better performance for this kernel.

Up to now, all performance improvements affected all cells of the domain. While the kernel runtime for updating $\mu$ is, up to measurement
error, equal in the complete domain, the $\phi$-kernel runtimes vary slightly due to branches in a routine that projects the $\phi$ values back into
the allowed simplex.
The following optimization introduces additional branches by skipping the evaluation of terms depending on the configuration of the 
current cell. This implementation is labeled as version ``with shortcuts'' in Figure~\ref{fig:sweep-optimizations} and was 
already included in the comparison of different vectorization strategies, shown above. 
These branches lead to different runtimes of the kernels
in different parts of the domain. The performance of the $\phi$-kernel is increased predominantly in liquid parts of the domain since there the
computation of the coupling term $\psi$ can be skipped entirely. The runtime of the $\mu$-kernel is improved especially in solid cells due to a simpler
calculation of the anti-trapping current in these cells.

All optimizations combined result in a total speedup of up to 80 depending on the architecture. 
This huge speedup compared to the original C implementation enables us to simulate sufficiently large domains in reasonable time to get physically meaningful results.

Since the relative speedup does not indicate to what extent the target architecture is utilized, we additionally investigated 
the absolute performance of our code. In the following, we focus on the singlenode performance of our optimized code, without the ``shortcut'' optimizations, since in this 
case the total number of executed floating point operations per cell can be determined exactly.

We show the performance analysis for the $\mu$-kernel on a SuperMUC node in detail, and shortly present the results for the $\phi$-kernel afterwards.
First, we use a roofline performance model to determine if the code is memory or compute bound \cite{rooflinePaper}. 
We measure the maximum attainable bandwidth using STREAM \cite{mccalpin1995memory} on one node, resulting in a bandwidth of approximately 80 GiB/s.
We assume that approximately half of the required data for one update can be held in cache, since in a symmetric D3C7 or D3C19 stencil
half of the data can be reused if one x-y-slice of both fields completely fits into cache. For a typical block size of $40^3$, one x-y-slice of all four fields fits into L2 cache.
Under this assumption, half of the required values are held in L2 cache and at most 680 Bytes have to be loaded from main memory to update one cell. 
For one cell update, 1384 floating point operations are required, resulting in a lower bound for the arithmetic intensity of approximately two floating point operations per byte. 
Taking only the memory bottleneck into account our code could achieve up to 126.3 MLUP/s on one SuperMUC node:

 \[  80 \, \frac{\text{GiB}}{\text{s}} \; : \; 680 \, \frac{\text{B}}{\text{LUP}} = 126.3 \, \text{MLUP/s}  \]

\begin{figure}%
  \begin{center} 
   \includegraphics[width=0.9\columnwidth]{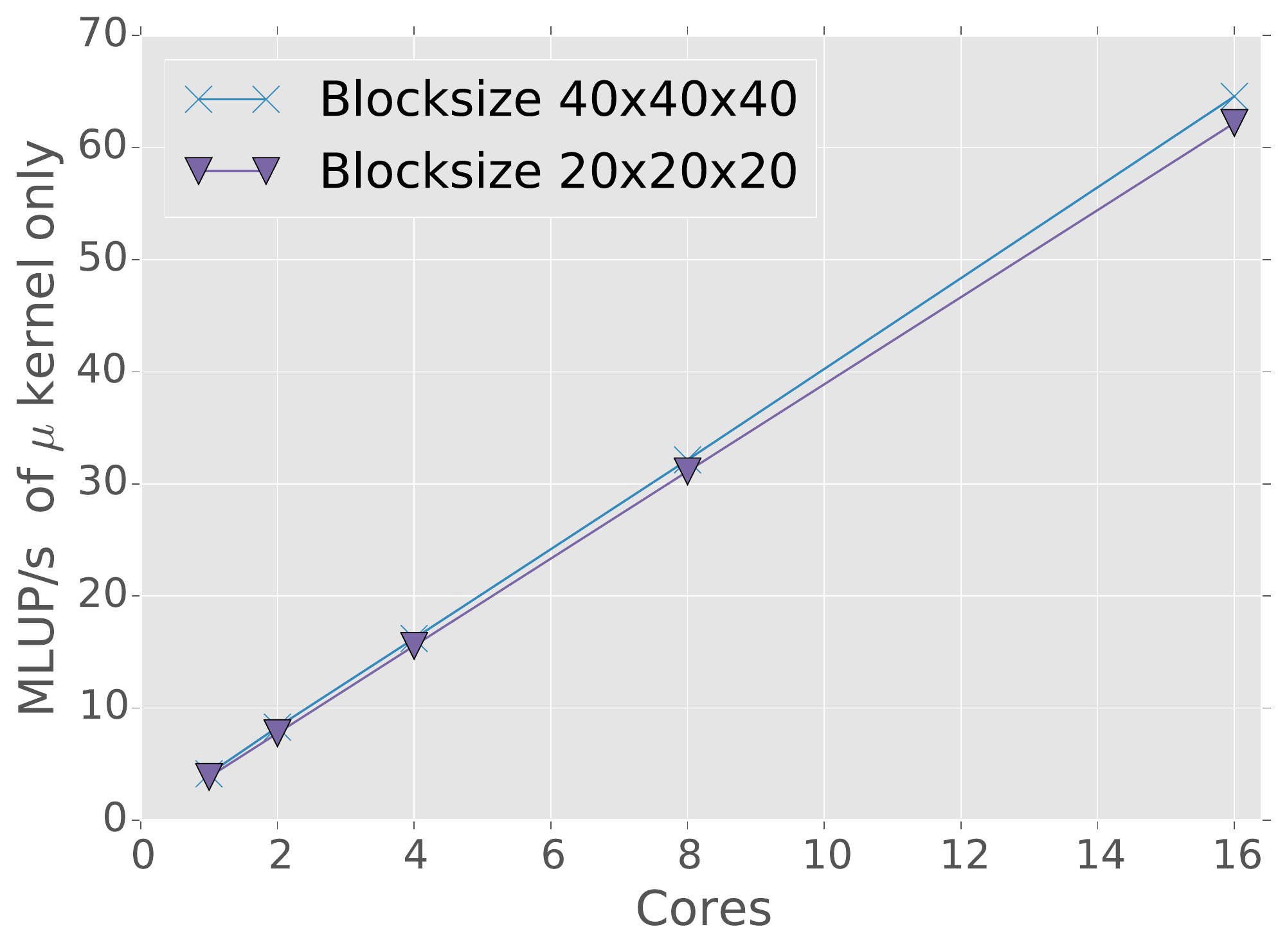}
  \end{center}
   \caption{Intranode Scaling of $\mu$-kernel without {\em shortcut} optimization on one SuperMUC node}
   \label{fig:mu_intranode_scaling}
\end{figure}

The benchmark results presented in Figure~\ref{fig:mu_intranode_scaling} show that our code does not hit the bandwidth limit of $126.3$ MLUP/s and therefore is not memory bound. Choosing
a smaller blocksize of $20^3$, which fits entirely into L3 cache, only changes the performance slightly. This is a further indication that the code is limited by in-core execution instead of memory bandwidth. 
To determine an upper bound for in-core execution we compare the attained FLOP/s rate with the maximal possible rate.
One SuperMUC core runs with a frequency of 2.7 GHz \cite{supermuc} and can perform 8 floating point operations per cycle, 
resulting in 21.6 GFLOP/s per core. The achieved 4.2 MLUP/s per core correspond to 5.8 GFLOP/s and therefore to 27\% peak performance of one core. 
The in-core execution time is therefore not limited by the total number of floating point operations. To understand why we cannot utilize the full computational capability
of one core, we employ the Intel Architecture Code Analyzer (IACA) \cite{iacaWebsite}. This tool inspects the assembly code of the kernel and statically predicts the estimated 
execution time on a given Intel architecture under the assumption that all data resides in L1D Cache.  
The IACA analysis shows that even though the code is fully vectorized, it can attain at most 43\% peak under ideal front-end, out-of-order engine, and memory hierarchy
conditions. This is caused predominantly by imbalance in the number of additions and multiplication as well as latencies for division operations.
The IACA report shows, that further, highly architecture dependent optimizations are possible, for example manually reordering addition and multiplication instructions.
These kind of optimizations would however require significant development effort due to the high complexity of the kernel of several thousand lines of code.
A similar analysis for the $\phi$-kernel shows that this kernel is also compute bound and attains approximately 21\% peak performance on one SuperMUC core.

\subsubsection{Scaling Results}

\begin{figure}%
  \centering
   \includegraphics[width=0.95\columnwidth]{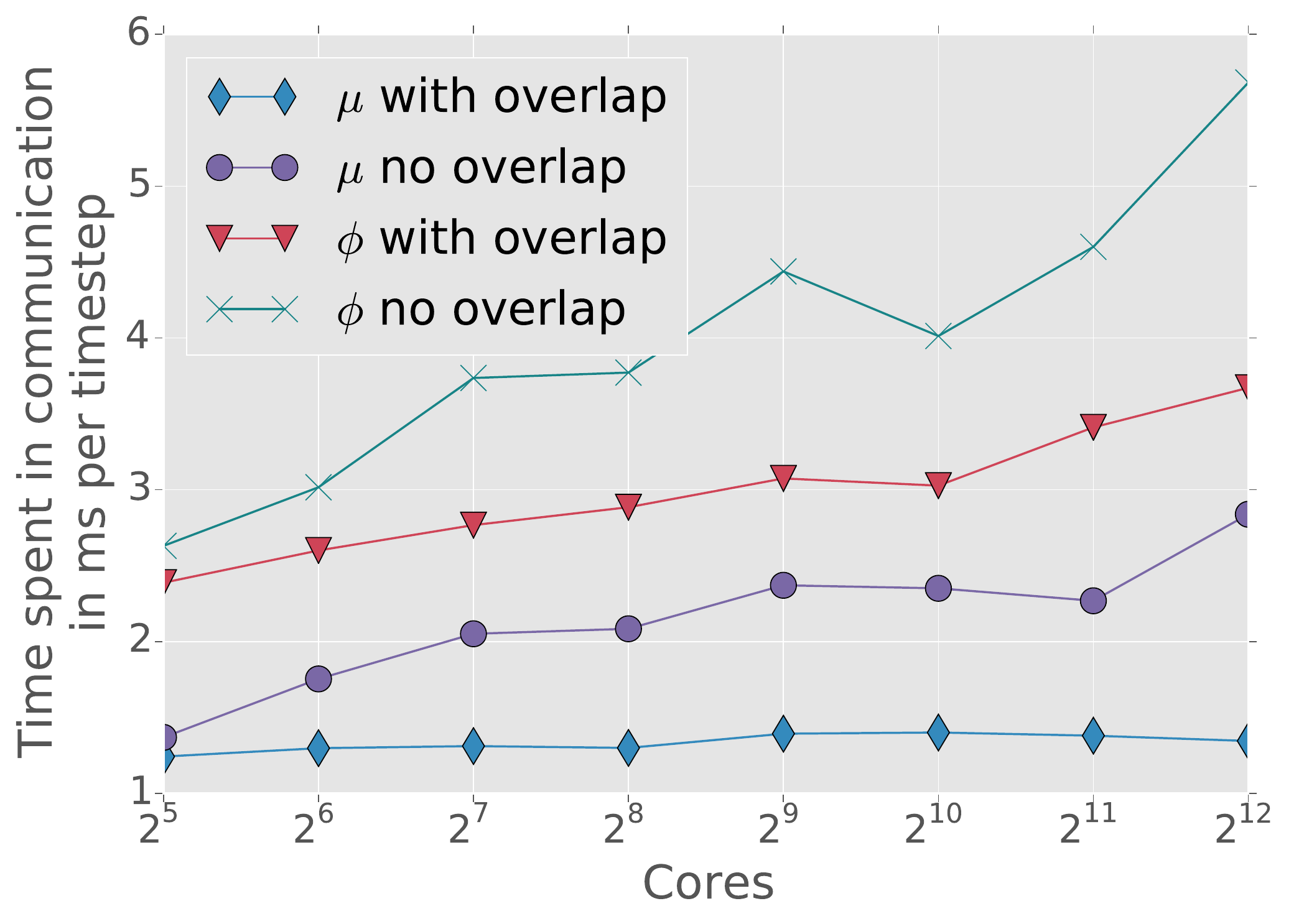}
  \caption{Time spent in communication, SuperMUC, blocksize $60^3$}
  \label{fig:comm_overlap}
\end{figure}

Before conducting scaling experiments we evaluate the effect of the presented communication hiding schemes.
All four possible combinations of hiding the communication of the $\phi$- and $\mu$-field were tested. 
Figure~\ref{fig:comm_overlap} shows the time spent in both communication routines. 
The amount of exchanged data is higher in the $\phi$-communication, thus the overall communciation times are higher in this case.
As expected, the effective communication times decrease for both fields when communication hiding is enabled. The remaining time in the communication
routines is spent for packing and unpacking messages which cannot be overlapped. 
Overlapping $\phi$-communication introduces additional overhead since the $\mu$-kernel has to be split up into two parts. 
As a consequence, the temperature dependent value computation, which is done once for each x-y-slice in the no-overlap case, 
has to be done twice when $\phi$ communication overlap is enabled. 
This overhead is much bigger than the benefit of communication hiding, thus the version with only $\mu$ communication hiding yields
the best overall performance.

\begin{figure*}[p]%
  \centering
   \includegraphics[width=0.32\textwidth]{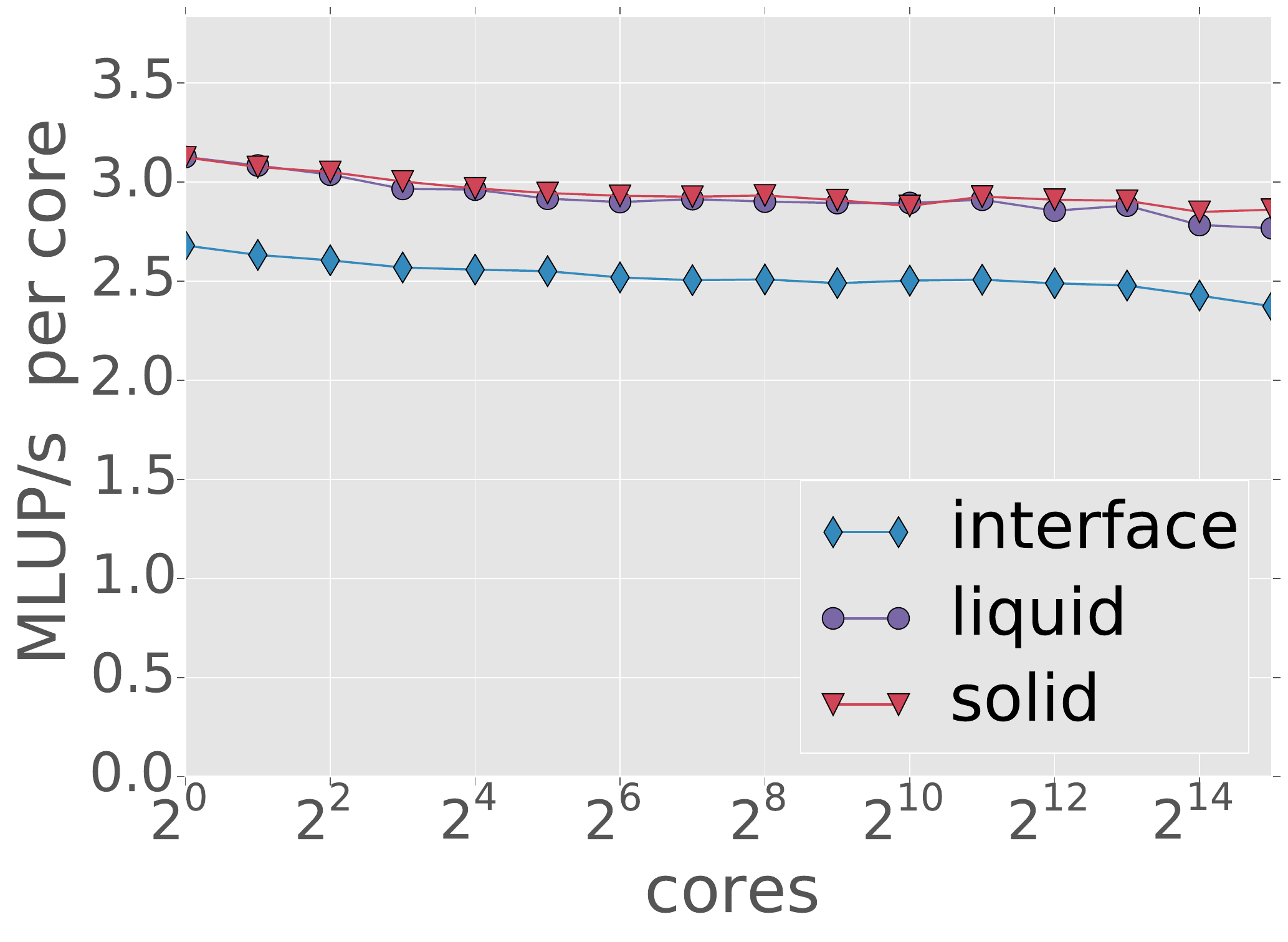}
   \includegraphics[width=0.32\textwidth]{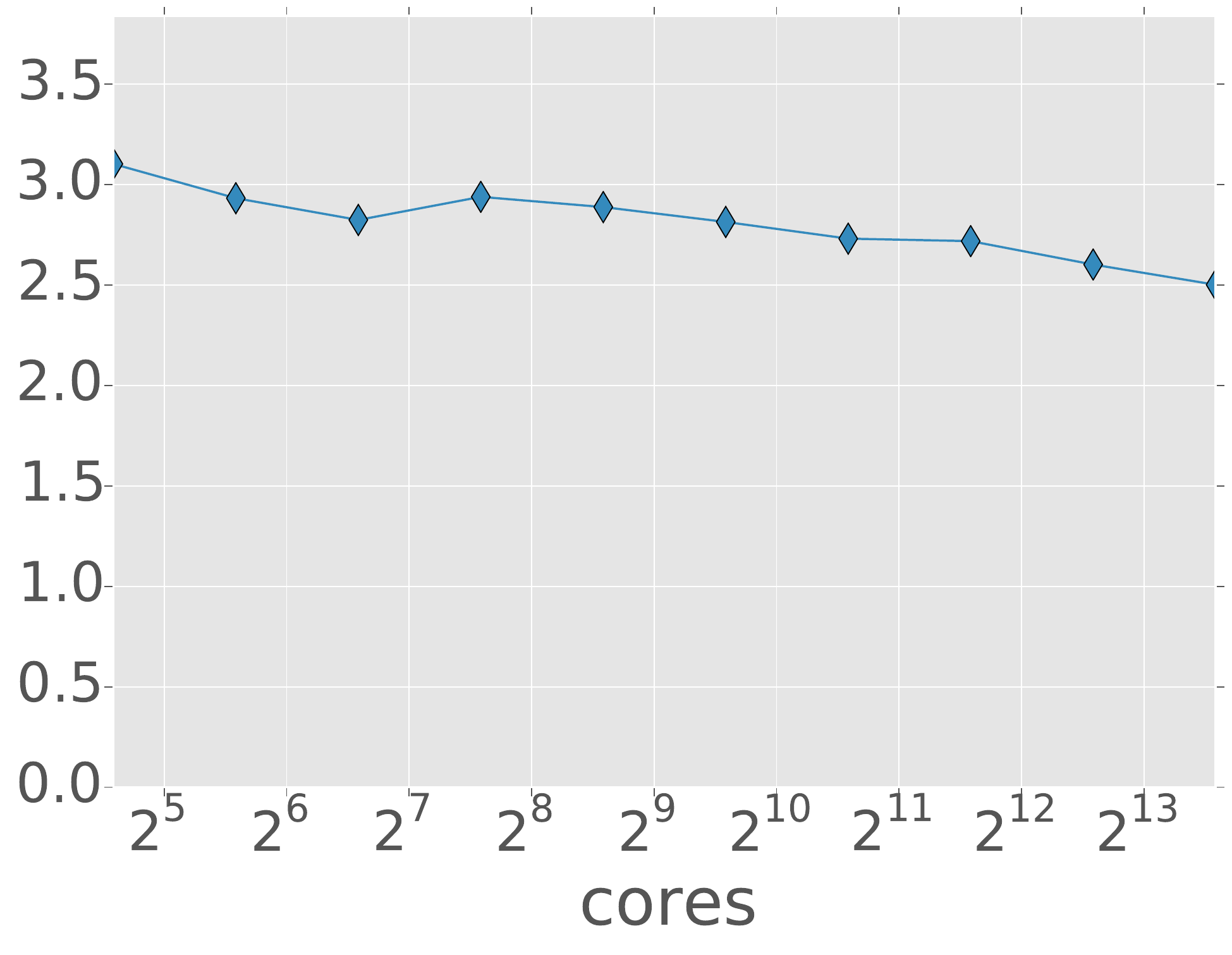}
   \includegraphics[width=0.32\textwidth]{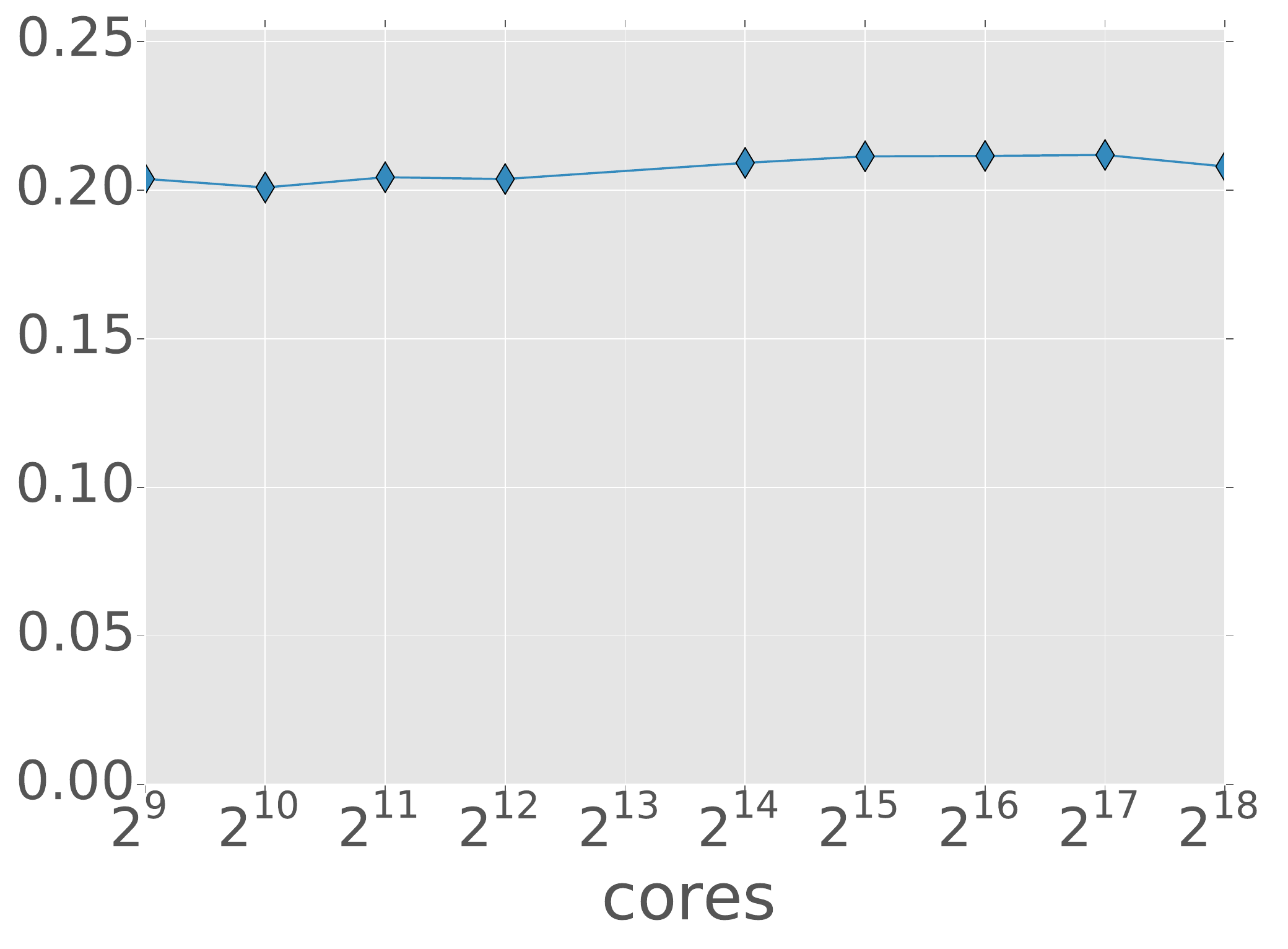}  
  \caption{Weak Scaling on SuperMUC(left), Hornet(middle) and JUQUEEN(right)}
  \label{fig:scaling}
\end{figure*}

In order to test the scalability of our implementation we execute weak scaling experiments on three of the largest German supercomputers.
Weak scaling scenarios, where the domain size and the process count are increased by the same factor,
are of practical relevance for this code, since the goal is to maximize the domain size while keeping 
the time to solution constant.
On SuperMUC, we scale three different scenarios up to 32,768 cores which were availabe to us. The full machine could not be utilized at the time of writing
due to installation of a hardware upgrade.
Figure~\ref{fig:scaling} shows that because of the ``shortcut'' optimization, we obtain slightly different runtimes
for the three scenarios, the ``interface'' scenario being the slowest due to higher workload in interface cells.
In production runs, where all of the three block compositions occur in the domain, the runtime is dominated by the interface blocks. 
We experimented with various load balancing techniques offered by the \walberla{} framework, which did, however, not decrease the
total runtime significantly, because the moving window technique makes it possible to simulate only the interface region,
such that, in production runs, most blocks have a composition similar to the ``interface'' benchmark.
For the scaling experiments on Hornet and JUQUEEN, we only used the ``interface'' scenario, which is representative for the 
performance achieved in production runs.

On SuperMUC and Hornet we placed one MPI process on each physical core. Simultaneous multithreading (SMT) did not improve the 
overall performance on these machines. 
On JUQUEEN however, 4-way SMT was employed to make full use of the in-order processing units. On this Blue Gene/Q 
machine we scaled up to 262,144 cores using 1,048,576 MPI processes.

\subsection{Simulation Results}

Due to the highly optimized code, we are able to study the evolution of microstructures in representative volume elements (RVE), which are required for a better understanding of the various pattern formations. 
The performed simulations allow us to study the evolution of microstructures, especially in three dimensions.

\begin{figure*}[htp]
\centering
  \subfigure[Simulation result]{
     \includegraphics[width=0.8\columnwidth]{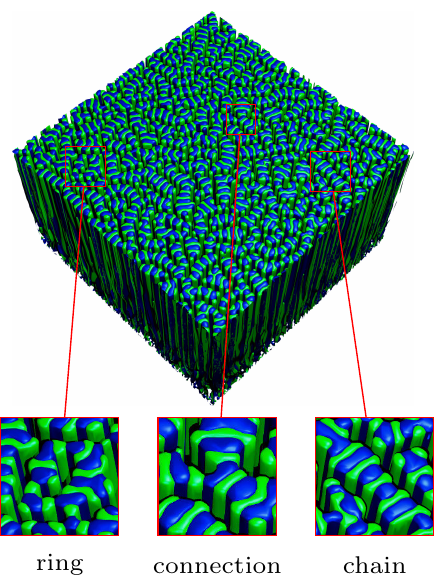}
     \label{fig:superlarge}
}
  \hspace{0.01\columnwidth}
  \subfigure[ tomography reconstruction of experiment from A.Dennstedt]{
     \includegraphics[width=0.8\columnwidth]{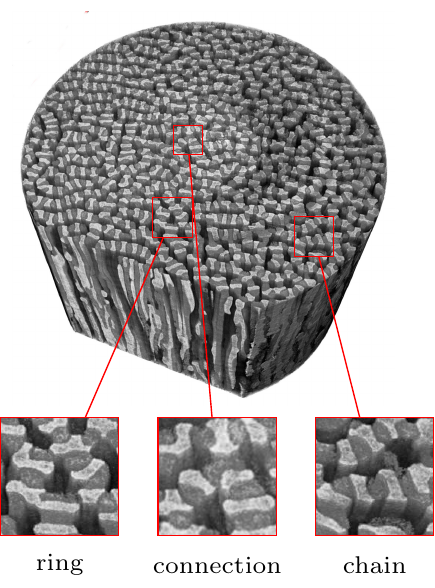}
     \label{fig:tomography}
}
 \caption{Three dimensional simulation and experimental results of directional solidification of the ternary eutectic system
 \systemAlAgCu{}}
  \label{fig:simexp}
\end{figure*}

\begin{figure*}[htp]
  \subfigure[Phase $Al_2Cu$]{
     \includegraphics[width=0.285\columnwidth,angle=270]{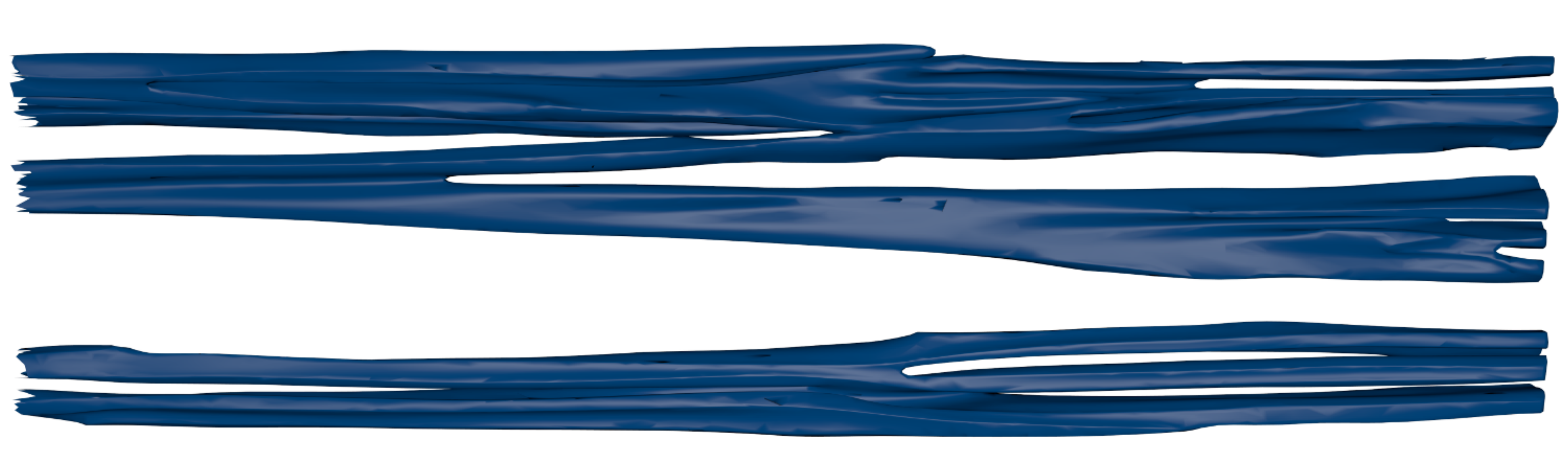}
     \label{fig:Al2Cu-lamellae}
}
  \hspace{0.005\columnwidth}
  \subfigure[Phase $Ag_2Al$]{
     \includegraphics[width=0.285\columnwidth,angle=270]{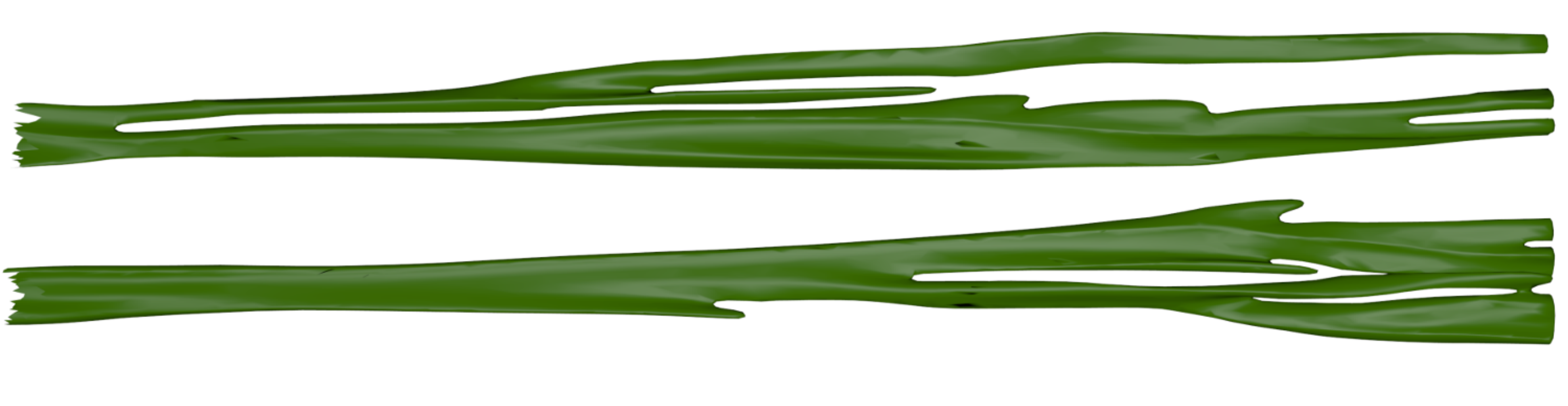}
          \label{fig:Ag_2Al-lamellae}
}
 \caption{Exempted lamellae from the simulation depicted in Figure~\ref{fig:simexp}. The lamellae grew from left to right.}
 \label{fig:lamellae}
\end{figure*}

Common experimental analyzing methods, like micrographs, only provide two dimensional information.
These micrographs can be used for a visual validation, but do not provide information on the three dimensional structure.
In comparison to experimental micrographs, we can show a good visual accordance with cross sections of our three dimensional simulations.
A simulation with $2420\times2420\times1474$ cells, simulated on the Hornet supercomputer, is depicted in Figure~\ref{fig:superlarge}.
In the experiment as well as the simulation, the phases arrange in similar patterns as chained brick-like structures that are connected or form ring-like structures, as shown in Figure~\ref{fig:simexp}.
Obtaining three dimensional information from experiments requires a large technical effort, using synchrotron tomography to resolve the different phases. 
A three dimensional reconstructed tomography result of directional solidification of the system \systemAlAgCu{} from A.~Dennstedt is depicted in Figure~\ref{fig:tomography}.
Even with these methods, only the final sample can be reconstructed, whereas the time evolution cannot be observed.
Using large-scale simulations, we found multiple microstructure characteristics, which could not be observed up to now due to small domain sizes.
Three dimensional simulations are crucial to capture all relevant physical phenomena: In two dimensional cross sections parallel to the growth front, the phases simple arrange in different lamellae as brick-like structures
while in three dimensions, various splits and merges of these lamellae can be observed. 
Single lamellae of the simulation in Figure~\ref{fig:simexp} are exempted in Figure~\ref{fig:lamellae}.
In Figure~\ref{fig:Al2Cu-lamellae}, two connected lamellae of the phase $Al_2Cu$ and in Figure~\ref{fig:Ag_2Al-lamellae} two lamellae of the phase $Ag_2Al$ are shown.  
The evolution of the microstructure, especially the splitting of lamellae and merging, is visible, and allows us to study the stability of different phase arrangements.
These phase arrangements influence the shape of the occurring phases and are of special importance to technical applications since the microstructure strongly influences the mechanical properties, e.g. the elastic and plastic deformation behavior.
Besides the good visual accordance of the large-scale simulations and two dimensional micrographs, also good agreement with the experimental 3D data is achieved.
A more detailed comparison is shown in \cite{hoetzer15} and a quantitative comparison using Principal Component Analysis on two-point correlation is in preparation.
In this publication, also the quantitative necessity of these large domain sizes will be presented.
In addition to the good visual agreement with the experiments, the simulations allow us to conduct parameter variations under well-defined conditions, 
which provide excellent ways to study the underlying physical mechanisms of the resulting microstructure formation.

\section{Conclusion}

In this work, we have developed and evaluated a massively parallel simulation code for a thermodynamic consistent phase-field model, based on the grand potential approach,
which runs efficiently on current supercomputing architectures. 
With this code, we simulated the directional solidification of ternary eutectics, including four phases and three chemical species.
Since big domain sizes are required to observe the formation of mircostructural patterns, a HPC implementation is crucial to get physical results for the considered scenario.
Starting from a general purpose and validated phase-field code of this complicated multiphase model for the solidification  of alloys, we developed a new, highly optimized implementation that can effectively utilize modern supercomputing architectures.
We applied optimizations on various levels to the highly complex stencil code, ranging from scenario dependent model simplifications to architecture specific optimizations like explicit SIMD vectorization. 
Systematic node level performance engineering resulted in a factor 80 speedup compared to the original code, as well as 25\% of peak performance on node level. 
Additionally, communication hiding techniques were applied to optimize the performance on system level, too. 
We have shown excellent scaling results of our code on three of the largest German supercomputers: SuperMUC, Hornet, and Juqueen. 
Only with this optimized implementation, it was possible to achieve the domain sizes necessary to gain microstructure patterns that are in good visual accordance with experimental data.
 
For future work, we plan to switch from the explicit Euler time stepping scheme to an implicit solver.
With the presented, optimized model, further parameter studies will be conducted.
Furthermore, support for a wider range of architectures (Xeon Phi, GPU) is in preparation.


\section{Acknowledgments}
We are grateful to the High Performance Computing Center Stuttgart, the Leibniz Rechenzentrum in Garching, and the J\"ulich Supercomputing Center for providing computational resources.
We also thank Dr. Anne Dennstedt from German Aerospace Center (DLR) in Cologne for providing the three dimensional synchrotron tomography reconstruction and the many fruitful discussions.


\pagebreak
\bibliographystyle{abbrv} 


\end{document}